\documentclass[letterpaper]{article} 

\usepackage{aaai25} 

\usepackage{times}  
\usepackage{helvet}  
\usepackage{courier}  
\usepackage[hyphens]{url}  
\usepackage{graphicx} 
\usepackage{natbib}  
\usepackage{caption} 
\usepackage{amsmath}
\usepackage{amsfonts}
\usepackage{algorithm}
\usepackage{algorithmic}
\usepackage{multirow}

\usepackage{newfloat}
\usepackage{listings}
\DeclareCaptionStyle{ruled}{labelfont=normalfont,labelsep=colon,strut=off} 
\floatstyle{ruled}
\newfloat{listing}{tb}{lst}{}
\floatname{listing}{Listing}
\usepackage{booktabs}

\title{Task-level Distributionally Robust Optimization for Large Language Model-based Dense Retrieval}
\author {
    Guangyuan Ma\textsuperscript{\rm 1,2}\footnote{
    Work done during internship at Langboat Technology.
},
    Yongliang Ma\textsuperscript{\rm 3},
    Xing Wu\textsuperscript{\rm 1,2},
    Zhenpeng Su\textsuperscript{\rm 1,2},
    Ming Zhou\textsuperscript{\rm 3},
    Songlin Hu\textsuperscript{\rm 1,2}\thanks{Corresponding author.}
}
\affiliations {
    \textsuperscript{\rm 1}Institute of Information Engineering, Chinese Academy of Sciences, Beijing, China\\
    \textsuperscript{\rm 2}School of Cyber Security, University of Chinese Academy of Sciences, Beijing, China\\
    \textsuperscript{\rm 3}Langboat Technology, Beijing, China\\
    \{maguangyuan,wuxing,suzhenpeng,husonglin\}@iie.ac.cn, \{mayongliang,zhouming\}@langboat.com
}

\begin{document}

\maketitle

\begin{abstract}
Large Language Model-based Dense Retrieval (LLM-DR) optimizes over numerous heterogeneous fine-tuning collections from different domains. However, the discussion about its training data distribution is still minimal. Previous studies rely on empirically assigned dataset choices or sampling ratios, which inevitably lead to sub-optimal retrieval performances. In this paper, we propose a new task-level Distributionally Robust Optimization (tDRO) algorithm for LLM-DR fine-tuning, targeted at improving the universal domain generalization ability by end-to-end reweighting the data distribution of each task. The tDRO parameterizes the domain weights and updates them with scaled domain gradients. The optimized weights are then transferred to the LLM-DR fine-tuning to train more robust retrievers. Experiments show optimal improvements in large-scale retrieval benchmarks and reduce up to 30\% dataset usage after applying our optimization algorithm with a series of different-sized LLM-DR models.
\end{abstract}

\begin{links}
    \link{Code}{https://github.com/ma787639046/tdro}
    \link{Datasets}{https://huggingface.co/tdro-llm}
\end{links}

\section{Introduction}
Dense retrieval \cite{karpukhin2020dpr} recalls relevant documents from large-sized candidate pools with the similarity search \cite{Stephen2016MIPS} of query-passage embeddings. The recent bloom of Large Language Model-based Dense Retrieval (LLM-DR) \cite{Wang2024mistral-E5, Meng2024SFR-Embedding, Muennighoff2024GritLM} promotes remarkable retrieval abilities with better foundation models \cite{Hugo2023LLaMA, Bai2024Qwen, Jiang2023Mistral-7B} and large-scale training collections \cite{Wang2024mistral-E5, Xiao2023BGE}. 

LLM-DR fine-tuning learns LLM-based retrieval models with heterogeneous training collections \cite{NilsReimers2021Sentence-Transformers-Training-Data} from multiple domains with different learning difficulties. The data distribution of training collections, i.e. a mixture of data with \textit{chosen datasets} or \textit{sampling ratio} on each dataset, significantly influences the general retrieval performances of dense retrievers \cite{Oren2019DRO, Meng2024SFR-Embedding}. However, the choice or sampling ratio of the training sets still relies heavily on intuitional assessments. It's hard to decide whether an empirical assigned data sampling ratio is optimal for the models to perform well, i.e. robust to all tasks. Nowadays, the robustness of data distributional optimization for LLM-DR is still very limited.

Distributionally Robust Optimization (DRO) \cite{Oren2019DRO, Sagawa2019GroupDRO, Piratla2022CGD-DRO} receives extensive discussions for battling unbalanced data composition. GroupDRO \cite{Sagawa2019GroupDRO}, the most popular algorithm for DRO, optimizes on the worst-case loss of the corresponding domain, which picks a domain with the highest loss at each training step and up-weights the loss of this domain. Although there was an attempt \cite{Yu2022COCO-DR} to utilize vanilla GroupDRO or variants of DRO algorithm \cite{Piratla2022CGD-DRO} for dense retrieval fine-tuning, the optimization is limited to a small BERT-based model over clustered groups of one single dataset, i.e. MS-MARCO \cite{tri2016msmarco}, which fails to generalize to LLM-DR fine-tuning with multiple heterogeneous training collections. It's profitable to solve the data distribution issue of LLM-DR in an end-to-end manner like DRO, but such a study is still left for further exploration.

\begin{figure}[!t]
\centering
\includegraphics[width=\linewidth]{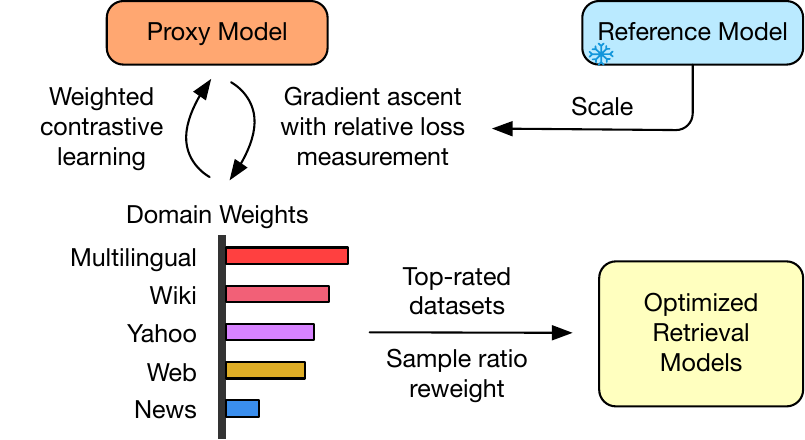}
\caption{
Task-level Distributionally Robust Optimization for Large Language Model-based Dense Retrieval.
}
\label{llmemb_overview}
\end{figure}

The existing DRO algorithms \cite{Oren2019DRO, Sagawa2019GroupDRO, Piratla2022CGD-DRO}, such as GroupDRO, are theoretically incompatible with LLM-DR fine-tuning due to \textit{different batch sampling strategies} and \textit{incommensurable loss scales}. Firstly, DRO requires all domain data mandatory in one batch for end-to-end comparisons. It dynamically reweights the training losses based on the worst-case group and derives the robust-optimized proxy model as the final model directly. 
However, the LLM-DR fine-tuning collects the heterogeneous sets in a homogeneous batching method \cite{Meng2024SFR-Embedding} during its fine-tuning, 
which means only one domain can be collected to the whole batch to ensure that in-batch negatives are sampled from the same task. What's worse, the heterogeneous collections used by LLM-DRs have significantly different loss scales. If directly applying the DRO algorithms \cite{Sagawa2019GroupDRO, Yu2022COCO-DR, Piratla2022CGD-DRO} to LLM-DRs, the models will always bias towards the domain with the highest training loss, i.e. worst case loss, which will hurt the fine-tuning process. As is shown in Table \ref{table_loss_compare_of_qwen500m}, the loss of Yahoo answers with a Qwen1.5-0.5B retriever is three times over MS-MARCO and five times over DuReader. Directly using worst-case loss will make the model always biased towards Yahoo, rather than MS-MARCO or DuReader.

\begin{table}[!t]
\centering
\small
    \begin{tabular}{l|c}
    \toprule  
    \textbf{Dataset} & \textbf{Loss} \\
    \midrule
    Yahoo answers (Title-Answer) & 3.9257 \\
    MS-MARCO & 1.3312 \\
    DuReader & 0.6925 \\
    \bottomrule
    \end{tabular}
\caption{
Comparison of loss scales for Yahoo answers (Title-Answer) \cite{Zhang2015YahooAnswers}, MS-MARCO \cite{tri2016msmarco}, and DuReader \cite{Qiu2022DuReader-Retrieval} datasets. The model used here is Qwen1.5-0.5B trained with uniform data sampling ratios for 1k steps.
}
\label{table_loss_compare_of_qwen500m}
\end{table}

To tackle the above optimization problems for LLM-DR, we develop a new task-level Distributionally Robust Optimization (tDRO) algorithm that aims at improving the general domain adaption ability across multiple \textit{tasks}\footnote{For simplicity, we treat each \textit{dataset} as a single \textit{domain} / \textit{group} / \textit{task} by following \cite{Sagawa2019GroupDRO}. We will mix the usage of these different expressions in the rest of our paper.}. 
Firstly, to coordinate with different batch sampling strategies, our algorithm separates the DRO optimization and LLM-DR fine-tuning as is presented in Figure \ref{llmemb_overview}.
Instead of directly learning a robust model \cite{Oren2019DRO, Sagawa2019GroupDRO}, such separation first learns domain weights with a proxy model via the DRO algorithm and then transfers learned domain weights to LLM-DR fine-tuning.
The proxy model is initialized from a small-sized LM, e.g. Qwen1.5-0.5B \cite{Bai2024Qwen}. It receives uniformly sampled training collections within an input batch, computes contrastive losses of each domain, and uses them as the gradients of domain weights. Then it transfers the learned weight or merely chooses the top-weighted datasets all LLM-DR model fine-tunings with different sizes. This separation shares several benefits: We can sample all domains within a batch in the tDRO stage and use task-homogeneous batching in the fine-tuning stage, which makes DRO work well while not hurting the final retrieval performances of LLM-DRs. Also, a small-sized LM can be used in tDRO optimizations to reduce computational overheads.

Secondly, as discussed above, the heterogeneous domains have different loss scales. To make a comparable measurement of domain running losses, we use a trained LLM-DR model, e.g. Qwen1.5-0.5B with uniform data sampling ratios, as the reference model and forward it with the same inputs. We compute the \textit{relative loss measurement} by dividing the proxy loss with reference loss. Intuitively, the relative loss measurement represents the improving headroom for the corresponding domains. Higher gradients will up-weight more corresponding domain weights. 

To verify the effectiveness of the tDRO algorithm, we conduct data distribution optimization on open-sourced sentence transformers training data \cite{NilsReimers2021Sentence-Transformers-Training-Data}.  
We test on three large-scale retrieval benchmarks to fully assess the universal retrieval abilities across different languages and domains, including multilingual MIRACL \cite{Zhang2023MIRACL}, cross-lingual MKQA \cite{Longpre2021MKQA}, and monolingual BeIR \cite{thakur2021beir}.
Experiments 
shows steady improvements with less dataset usage after applying tDRO optimization.

\section{Algorithm}

\subsection{Problem Statement}
Assume the training collections $D^{train}$ of LLM-DR fine-tuning are composed of $k$ individual datasets, each of them is assigned with a domain weight $\alpha$, representing a choice of probability distribution $P_{\alpha}$ over joint training collections:

\begin{equation}
\begin{aligned}
    P_{\alpha} = \sum_{g=1}^{k} \alpha_{g}\mathrm{U}(D_{g}^{train}), 
    s.t. \sum_{g=1}^{k}\alpha_{g} = 1.
\end{aligned}
\end{equation}

\noindent $\mathrm{U}$ is a uniform distributional sampling function. And $\alpha_{g}\mathrm{U}(D_{g}^{train})$ means sampling from such distribution with weight $\alpha_{g}$ for group $g$, which is also called $\alpha$-covered probability distribution \cite{Oren2019DRO}. The goal of LLM-DR data distributional optimization is to find an optimal distribution $P_{\alpha}$, or a choice of weights $\alpha$ specifically, enabling the model to perform well on all downstream tasks $D^{test}$. Note that downstream tasks $D^{test}$ are not necessarily identical to the fine-tuning sets $D^{train}$.

\begin{figure*}[t!]
\centering
\includegraphics[width=14cm]{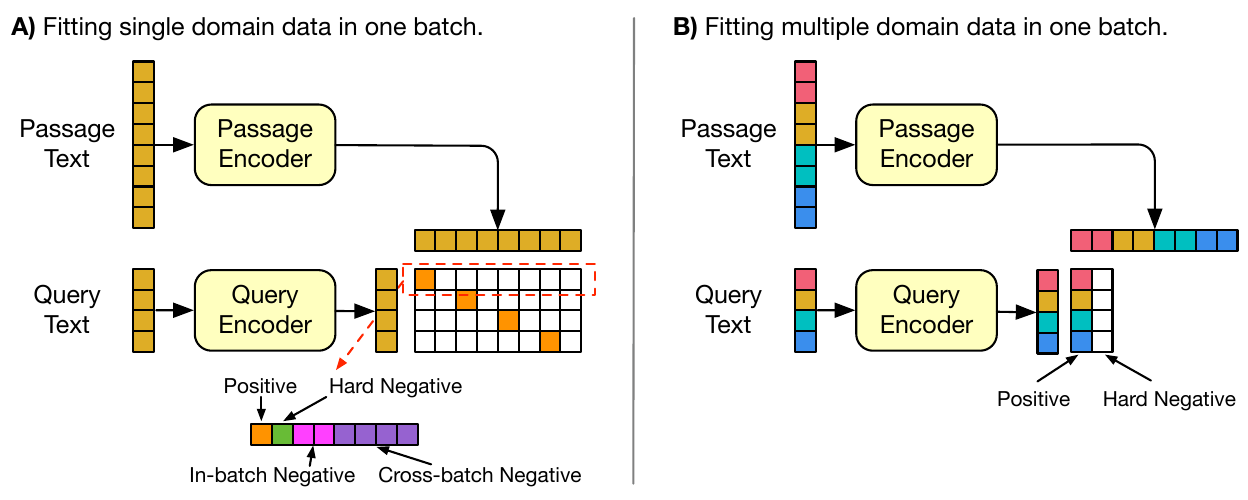}
\caption{
Different batch sampling strategies and negative types for A) LLM-DR Fine-tuning and B) Distributionally Robust Optimization.
}
\label{llmemb_pipeline}
\end{figure*}

\subsection{Task-level Distributionally Robust Optimization}
The task-level Distributionally Robust Optimization (tDRO) algorithm parameterizes domain weights $\alpha$ and tries to learn the best choice of $\alpha$ for all tasks in an end-to-end manner. The tDRO holds a basic assumption for solving the data distributional optimization of LLM-DR: In a scenario with multiple independent and identically distributed (iid) task collections, if a model is robust to the training phase, then it will be robust to most of the test sets. Thus like most DRO algorithms \cite{Oren2019DRO, Sagawa2019GroupDRO, Piratla2022CGD-DRO}, tDRO operates on the robustness of training collections for universal adaption abilities. The whole optimization pipeline includes a separate tDRO stage and LLM-DR fine-tuning stage.

\subsubsection{InfoNCE Loss}
tDRO treats each dataset as a task (or domain) at minimal granularity and trains a proxy model with parameters $\theta_{proxy}$ to adaptively compare and update the domain weights at the task level. At each training step $t$, a batch of training data with domain sampling probabilities $1/k$ is forwarded through the proxy model. By extracting hidden states from the last token position, each batch item comprises a query representation $q$, a positive document representation $d^+$, and several hard negative ($HN$) document representations $d^{-}_{HN}$. InfoNCE loss \cite{Oord2018InfoNCE}, i.e. contrastive loss, is used to calculate losses of the proxy model for each group:

\begin{equation}
    \mathcal{L}^{proxy}=-\log \frac{e^{q \cdot d^{+}/\tau}}{e^{q \cdot d^{+}/\tau} + \sum e^{q \cdot d^{-}_{HN}/\tau}},
    \label{formula_proxy_loss}
\end{equation}

\noindent where $\tau$ is a contrastive temperature. In-batch negative sampling is not an option for tDRO because different tasks could induce false negatives and reduce negative qualities. 

\begin{table*}[!ht]
\centering
\setlength{\tabcolsep}{1mm}
\small
    \begin{tabular}{l|c|c|c}
    \toprule
        \textbf{Dataset} & \textbf{Language} & \textbf{Category} & \textbf{Deduped Size} \\
        \midrule
        agnews \cite{Corso2005agnews} & English & News & 1,157,745 \\
        AllNLI \cite{bowman2015SNLI, Williams201MNLI} & English & NLI & 277,230 \\
        altlex \cite{Hidey2016atlex} & English & Wikipedia Pair & 112,696 \\
        amazon\_review\_2018 \cite{Ni2019AmazonReview2018} & English & Amazon & 999,999 \\
        cnn\_dailymail \cite{see2017CNN-Daily-Mail} & English & News & 311,971 \\
        codesearchnet \cite{Husain2019codesearchnet} & English & Github & 1,375,067 \\
        dureader \cite{Liu2021baidu_search} & Chinese & Web Collections & 86,395 \\
        eli5\_question\_answer \cite{Fan2019ELI5} & English & Reddit & 325,390 \\
        gooaq\_pairs \cite{Khashabi2021GooAQ} & English & Web Collections & 3,012,347 \\
        hotpotqa \cite{Yang2018HotpotQA} & English & Wikipedia QA & 85,000 \\
        medmcqa \cite{Pal2022MedMCQA} & English & Medical & 160,865 \\
        miracl \cite{Zhang2023MIRACL} & 16 languages & Multilingual Wikipedia & 32,405 \\
        mr\_tydi\_combined \cite{Zhang2021MrTyDi} & 11 languages & Multilingual Wikipedia & 48,475 \\
        msmarco \cite{tri2016msmarco} & English & Web Collections & 502,854 \\
        nq \cite{tom2019nq} & English & Wikipedia QA & 58,800 \\
        quora\_duplicates\_triplets \cite{quora-pairs} & English & Forum Duplicates & 97,011 \\
        searchQA\_top5\_snippets \cite{Dunn2017SearchQA} & English & Web Collections & 117,219 \\
        sentence-compression \cite{Filippova2013SentenceCompression} & English & News & 180,000 \\
        SimpleWiki \cite{Coster2011SimpleWiki} & English & Wikipedia Pair & 102,225 \\
        squad\_pairs \cite{Rajpurkar2016SQuAD} & English & Wikipedia QA & 87,595 \\
        stackexchange\_duplicates\_title-body \cite{StackExchange-Duplicates} & English & Forum Duplicates & 250,516 \\
        t2ranking \cite{Xie2023T2Ranking} & Chinese & Web Collections & 200,376 \\
        trivia \cite{Mandar2017TriviaQA} & English & Wikipedia QA & 60,370 \\
        xsum \cite{NarayanCL2018xsum} & English & News & 226,711 \\
        yahoo\_answers\_title\_answer \cite{Zhang2015YahooAnswers} & English & Yahoo & 1,198,018 \\
    \bottomrule
    \end{tabular}
\caption{Training datasets information. Note that strict deduplication is performed on all training sets with Simhash \cite{Manku2007SimHash} to ensure no overlap between training and testing sets.}
\label{table_datasets}
\end{table*}

\subsubsection{Optimization Objective}
tDRO learns domain weights $\alpha$ with a dual-optimization objective, which minimizes the supremum of the $\alpha$-weighted sum of group loss measurements. Such an objective ensures universal robustness by lowering the upper bound of the worst groups:

\begin{equation}
\label{equation_dro_minimax}
\begin{aligned}
    \mathop{\mathrm{min}}_{\theta} \mathop{\mathrm{sup}}_{\alpha} \sum_{g=1}^{k} \alpha_{g}\mathcal{M}_{g}(\theta_{proxy}; q, d^{+}, d^{-}),
\end{aligned}
\end{equation}

\noindent where \textit{group loss measurement} $\mathcal{M}_{g}$ is a scalar corresponding to averaged losses, representing the improving headrooms for each group. Following the previous DRO framework \cite{Sagawa2019GroupDRO}, the above object is optimized by interleaving the updates of weights $\alpha$ and proxy model parameters $\theta$.

\subsubsection{Weights Update}
For weights update, tDRO optimizes the above object by up-weighting the corresponding domain weights with exponential gradient ascent:

\begin{equation}
\label{equation_dro_alpha_update}
\begin{aligned}
    &\alpha_{g}^{t} = \alpha_{g}^{t-1}e^{\eta_{\alpha}\mathcal{M}_{g}}, \\
\end{aligned}
\end{equation}

\noindent where $\eta_{\alpha}$ is the learning rate for domain weights.
As a common optimation practice, gradient normalization is used on the above gradients to ensure stable weight updates. Intuitively, a higher loss measurement induces more up-weighting of the corresponding group. 
After the update of $\alpha$, a re-normalization is performed to ensure $\sum_{g=1}^{k}\alpha_{g} = 1$.

\subsubsection{Relative Loss Measurement}
The key component of tDRO is how to derive a balanced and comparable loss measurement $\mathcal{M}_{g}$. GroupDRO directly uses the average group loss $\mathbb{E}(\mathcal{L}_{g}^{proxy})$ as the loss measurement, where $\mathbb{E}$ is the arithmetic mean function. However, as presented in Table \ref{table_loss_compare_of_qwen500m}, the averaged contrastive losses of each group are not comparable. Directly using the average group loss will always make the proxy model biased toward one group with the highest loss. To solve the above issue, we introduce a trained reference model $ref$, forward it with the same inputs, compute reference loss as Formula (\ref{formula_proxy_loss}), and \textit{divide} the proxy loss with reference loss to dynamically scale the average group loss. This design is called \textit{relative loss measurement}.

\begin{equation}
    \mathcal{M}_{g} = \mathbb{E}(\mathcal{L}_{g}^{proxy}) / \mathbb{E}(\mathcal{L}_{g}^{ref}).
\end{equation}

The reference model is frozen and will not be updated during the tDRO stage. In our implementation, we initialize the reference model with Qwen1.5-0.5B \cite{Bai2024Qwen} and train with uniform sampling weights on all training sets. The training setting of the reference model follows the LLM-DR fine-tuning recipe, which will be described later.

\subsubsection{Proxy Model Update}
After updating domain weights, the proxy model is updated with $\alpha$-weighted contrastive loss.

\begin{equation}
\label{equation_groupdro_theta_update}
    \theta^{t} = \theta^{t-1} - \alpha_{g}\eta_{\theta}\nabla_{\theta},
\end{equation}

\noindent where $\eta_{\theta}$ is the learning rate of the proxy model, and $\nabla_{\theta}$ is the gradient of proxy model obtained by backpropagation. The AdamW optimizer can be used for fast convergence.

\subsection{LLM-DR Fine-tuning}
LLM-DR fine-tuning also uses contrastive loss to pull positive representations together and push negative representations away. But it uses 
completely different batch sampling strategies and negative types from tDRO, which is one of the reasons that previous DRO algorithms like GroupDRO are theoretically incompatible with LLM-DR fine-tuning.

\subsubsection{LLM-DR Batching Strategy}
As is shown in Figure \ref{llmemb_pipeline}A, 
LLM-DR fine-tuning fits the data from one single domain in each batch to ensure the quality of in-batch negatives. This batching strategy is also called task-homogeneous batching \cite{Meng2024SFR-Embedding}. As is a common practice, LLM-DR trains with large batch sizes, e.g. 2048 in our implementation, and three types of negatives, including hard negatives ($HN$), in-batch negatives ($IBN$), and cross-batch negatives ($CBN$). Large batch size enables contrastive learning of LLM-DR using more negatives, especially in-batch and cross-batch negatives. Hard negatives are provided by individual datasets, which are mined from existing retrieval systems like BM25 \cite{Robertson1994OkapiBM25} or dense retrievers \cite{karpukhin2020dpr}. In-batch negatives are samples from other data items within a batch. 
Cross-batch negatives are samples gathered from other GPU batches. Overall, the contrastive loss ($\mathcal{L}^{CL}$) for LLM-DR is formulated as follows.


\begin{equation}
    \mathcal{L}^{CL}=-\log \frac{e^{q \cdot d^{+}/\tau}}{e^{q \cdot d^{+}/\tau} + \sum e^{q \cdot \{d^{-}_{HN};d^{-}_{IBN};d^{-}_{CBN}\}/\tau}}.
\end{equation}

However, tDRO compares and updates domain weights in an end-to-end manner, thus it requires fitting all domain data in one batch. 
As displayed in Figure \ref{llmemb_pipeline}B, 
The training batch is composed of all domains, while the in-batch/cross-batch negatives are not used in tDRO.

\begin{table*}[!ht]
\centering
\small
\setlength{\tabcolsep}{1mm}
    \begin{tabular}{l|ccccc}
    \toprule
        \textbf{Benchmark} \textit{(\# Dataset)} & \multicolumn{2}{c}{\textbf{MIRACL} \textit{(18)}} & \multicolumn{2}{c}{\textbf{MKQA} \textit{(25)}} & \textbf{BeIR} \textit{(15)} \\
        \midrule
        \textbf{Metric} & nDCG@10 & Recall@100 & Accuacy@10 & Accuacy@100 & nDCG@10 \\
        
        \toprule
        \multicolumn{6}{l}{\textbf{Uniform Sampling Baselines}}  \\
        \midrule
        Qwen-0.5B & 45.8  & 80.5  & 43.1  & 61.3  & 47.5  \\
        Qwen-1.8B & 50.9  & 84.7  & 45.0  & 64.0  & 48.8  \\
        Qwen-4B & 55.9  & 88.7  & 53.7  & 70.2  & 51.8  \\
        Qwen-7B & 59.6  & 90.6  & 58.7  & 73.6  & 52.3  \\
        Mistral-7B & 61.3  & 91.6  & 59.8  & 72.8  & 55.2  \\
        Llama3-8B & 64.1  & 92.8  & 64.0  & 75.8  & 55.0  \\
        
        \toprule
        \multicolumn{6}{l}{\textbf{tDRO - Dataset Selection Top-70\%}} \\
        \midrule
        Qwen-0.5B & 48.7\textsuperscript{*} (+2.9)  & 82.1\textsuperscript{*} (+1.6) & 45.4\textsuperscript{*} (+2.3) & 62.3\textsuperscript{*} (+1.0) & 48.9\textsuperscript{*} (+1.4) \\
        Qwen-1.8B & 54.1\textsuperscript{*} (+3.2) & 86.6\textsuperscript{*} (+1.9) & 48.6\textsuperscript{*} (+3.6) & 65.6\textsuperscript{*} (+1.6) & 50.2\textsuperscript{*} (+1.4) \\
        Qwen-4B & 58.6\textsuperscript{*} (+2.7) & 90.0\textsuperscript{*} (+1.3) & 57.0\textsuperscript{*} (+3.3) & 71.4\textsuperscript{*} (+1.2) & 52.6\textsuperscript{*} (+0.8) \\
        Qwen-7B & 61.6\textsuperscript{*} (+2.0) & 91.4\textsuperscript{*} (+0.8) & 59.9\textsuperscript{*} (+1.2) & 73.8 (+0.2) & 53.3\textsuperscript{*} (+1.0) \\
        Mistral-7B & 63.8\textsuperscript{*} (+2.5) & 92.4\textsuperscript{*} (+0.8) & 62.5\textsuperscript{*} (+2.7) & 73.8\textsuperscript{*} (+1.0) & 55.2 (+0.0) \\
        Llama3-8B & \textbf{66.4}\textsuperscript{*} (+2.3) & \textbf{93.5}\textsuperscript{*} (+0.7) & 66.0\textsuperscript{*} (+2.0) & 76.4\textsuperscript{*} (+0.6) & 55.1 (+0.1) \\
        \midrule
        \textit{Avg Gains} & \textbf{+2.6} & \textbf{+1.2} & +2.5 & +0.9 & \textbf{+0.8} \\
        
        \toprule
        \multicolumn{6}{l}{\textbf{tDRO - Sample Ratio Reweighting}} \\
        \midrule
        Qwen-0.5B & 49.1\textsuperscript{*} (+3.3) & 82.7\textsuperscript{*} (+2.2) & 45.5\textsuperscript{*} (+2.4) & 62.2\textsuperscript{*} (+0.9) & 48.3\textsuperscript{*} (+0.8) \\
        Qwen-1.8B & 53.6\textsuperscript{*} (+2.7) & 86.5\textsuperscript{*} (+1.8) & 50.5\textsuperscript{*} (+5.5) & 66.8\textsuperscript{*} (+2.8) & 49.7\textsuperscript{*} (+0.9) \\
        Qwen-4B & 58.4\textsuperscript{*} (+2.5) & 90.0\textsuperscript{*} (+1.3) & 57.8\textsuperscript{*} (+4.1) & 72.0\textsuperscript{*} (+1.8) & 51.9 (+0.1) \\
        Qwen-7B & 61.0\textsuperscript{*} (+1.4) & 91.1 (+0.5) & 59.8\textsuperscript{*} (+1.1) & 73.6 (+0.0) & 52.4 (+0.1) \\
        Mistral-7B & 63.4\textsuperscript{*} (+2.1) & 92.4\textsuperscript{*} (+0.8) & 62.8\textsuperscript{*} (+3.0) & 74.0\textsuperscript{*} (+1.2) & \textbf{55.4} (+0.2) \\
        Llama3-8B & 66.3\textsuperscript{*} (+2.2) & 93.4\textsuperscript{*} (+0.6) & \textbf{67.0}\textsuperscript{*} (+3.0) & \textbf{76.8}\textsuperscript{*} (+1.0) & 55.0 (+0.0) \\
        \midrule
        \textit{Avg Gains} & +2.4 & \textbf{+1.2} & \textbf{+3.2} & \textbf{+1.3} & +0.4 \\
    \bottomrule
    \end{tabular}
\caption{
Retrieval performances and corresponding gains of tDRO algorithm on MIRACL dev, MKQA test, and BeIR test benchmarks. The highest retrieval scores and average performance gains are marked as \textbf{bold}. \textsuperscript{*}Significant improvements (p $\leq$ 0.01) over the corresponding baseline. MS-MARCO in BeIR uses the dev split because there is no public test label.
}
\label{table_main_results}
\end{table*}

\begin{figure*}[t!]
\centering
\includegraphics[width=\linewidth]{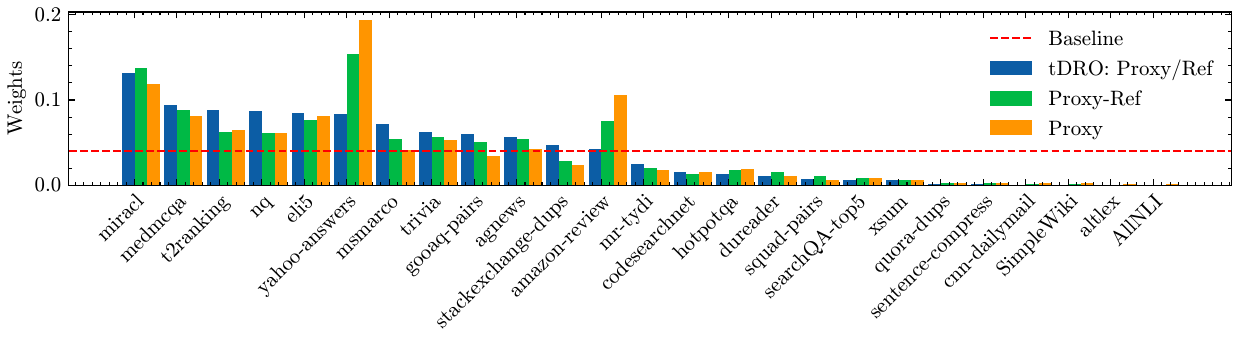}
\caption{
Weights comparison among baseline, tDRO, and other loss measurement designs.
}
\label{fig_weights}
\end{figure*}

\section{Experiments}

\subsection{Experiment Settings}

\subsubsection{Datasets}
A total of 25 individual datasets are used in our experiments, covering categories of Wikipedia, web collection, news, GitHub, yahoo, etc. Most of them are directly taken from sentence transformers training data \cite{NilsReimers2021Sentence-Transformers-Training-Data}. BGE-large-en-1.5 \cite{Xiao2023BGE} is used to mine negatives if the original datasets (several English datasets) have no negatives provided. Several popular multilingual datasets are also included in the training sets, including MIRACL \cite{Zhang2023MIRACL} and Mr.Tydi \cite{Zhang2021MrTyDi}. 
All information about datasets is listed in Table \ref{table_datasets}.

\subsubsection{Settings}
For the tDRO algorithm, both the proxy and reference models are initialized from Qwen1.5-0.5B \cite{Bai2024Qwen} for computational efficiency. tDRO is performed with a total batch size of 2048, query \& document maximum sequence lengths of 128 \& 512, a proxy model learning rate $\eta_{\theta}$ of 1e-4, contrastive temperature $\tau$ of 0.002, weights learning rate $\eta_{\alpha}$ of 2e-2, and seed of 42.
The weights from the tDRO stage are directly transferred to LLM-DR fine-tuning. Two weight transfer strategies are utilized: 
\begin{enumerate}
    \item \textbf{Top-rated dataset selection}: Use the \textit{Top-rated} datasets and discard the datasets with lower weights. This helps reduce dataset usage.
    \item \textbf{Sample Ratio Reweighting}: Directly use the weights to \textit{reweight} the sample ratios of datasets. 
\end{enumerate}

We conduct weights transfer on Qwen-1.5  0.5B, 1.8B, 4B, 7B \cite{Bai2024Qwen}, Mistral-0.1-7B \cite{Jiang2023Mistral-7B} and Llama3-8B \cite{Hugo2023LLaMA} for LLM-DR fine-tuning. Contrastive learning is performed with the same batch size, sequence lengths, model learning rate (1e-4), and contrastive temperature as stated before. Gradient cache \cite{Gao2021GradCache}, flash attention 2 \cite{TriDao2023FlashAttention2}, full-shard data parallel (FSDP), activation checkpointing and low-rank adapter (LoRA) \cite{Hu2022LoRA} with $r=8, \alpha=32$ and dropout ratio of 0.1 are used to reduce GPU memory usage. Following the previous work \cite{Wang2024mistral-E5, Su2023Instructor}, prompt instructions are added on the query side for multi-tasks during training and evaluation. 
All trainings are performed on 8 NVIDIA H800 GPUs with 4.5 hours for tDRO and less than 10 hours for all LLM-DR fine-tunings.

\begin{table}[!t]
\small
\setlength{\tabcolsep}{1mm}
\centering
    \begin{tabular}{l|c|ccc}
    \toprule
        \textbf{Benchmark} & \textbf{Enhance} & \textbf{MIRACL} & \textbf{MKQA} & \textbf{BeIR} \\
        \midrule
        \textbf{Metric $\rightarrow$} & \textbf{Special} & \textbf{nDCG@10} & \textbf{Acc@100} & \textbf{nDCG@10} \\
        \textbf{Models} $\downarrow$ & \textbf{Pre-train} & ~ & ~ & ~ \\
        \midrule
        BM25 & ~ & 31.9  & 39.9  & 41.7  \\
        mContriever & \checkmark & 43.1  & 67.9  & 46.0  \\
        mE5-large-inst & \checkmark & 64.6  & 70.2  & 52.3  \\
        E5-Mistral\textsuperscript{*} & ~ & 62.2  & 68.6  & 52.7  \\
        \midrule
        \multicolumn{5}{l}{\textbf{tDRO + LLM-DR Llama3-8B}} \\
        \midrule
        Top-70\% & ~ & \textbf{66.4}  & 76.4  & \textbf{55.1}  \\
        Reweighting & ~ & 66.3  & \textbf{76.8}  & 55.0 \\
    \bottomrule
    \end{tabular}
\caption{
Retrieval scores among state-of-the-art retrievers. \textsuperscript{*}We take the released BeIR scores without GPT data for fair comparisons.
}
\label{table_sota_scores}
\end{table}

\subsubsection{Baselines and Benchmarks}
To compare the performance changes after tDRO optimization, we choose LLM-DR with \textit{uniform sampling weights}\footnote{A total of 25 datasets are used in our experiments, thus the sampling weight for each dataset is $1/25=0.04$} as baselines. All other settings are kept the same. Note that results from some recent state-of-the-art retrievers, including BM25 \cite{Robertson1994OkapiBM25}, mContriever/Contriever \cite{Izacard2021contriever}, E5-Mistral-7b \cite{Wang2024mistral-E5}, and Multilingual-e5-large-instruct \cite{Wang2024mE5} are also added to our results. But we DO NOT seek to compare with them. Some of them utilize multiple training enhancements, such as data synthesis with ChatGPT and special pre-trainings \cite{wang2022simlm, wu2023contextual, Ma2024drop_decoder} for retrieval, which is an unfair comparison to our experiments and out of the research scope for our paper.

MIRACL \cite{Zhang2023MIRACL}, MKQA \cite{Longpre2021MKQA}, and BeIR \cite{thakur2021beir} are used as main retrieval benchmarks. MIRACL is a huge multilingual retrieval benchmark across 18 languages with 13K queries and 90M passages. Cross-lingual benchmark MKQA holds 25 different languages of parallel Wikipedia queries. Following \cite{Chen2024BGE-M3-Embedding}, the cross-lingual retrieval is performed by using queries (6.6k for each) of different languages to recall relevant English Wikipedia with 2.7M passages\footnote{https://huggingface.co/datasets/Tevatron/wikipedia-nq-corpus}. BeIR is a heterogeneous English retrieval benchmark with 15 public datasets and 34M documents, covering a wide range of retrieval domains and text symmetries. MIRACL is reported with nDCG@10 and Recall@100. MKQA is reported with Accuacy@\{10, 100\} (Acc). Examing performances at threshold 10 assess the top-ranking ability and threshold 100 for recalling capacity at a larger window. BeIR is reported with nDCG@10 by following the original paper \cite{thakur2021beir}. The whole evaluation takes around 30 hours with 8 NVIDIA H800 GPUs for the largest LLM-DR retriever, Llama3-8B.

\begin{table}[!t]
\small
\setlength{\tabcolsep}{1mm}
    \centering
    \begin{tabular}{l|c|c|c}
    \toprule
        ~ & \textbf{BEIR} & \textbf{MIRACL} & \textbf{MKQA} \\
        \midrule
        \textbf{Metric} & nDCG@10 & nDCG@10 & Acc@100 \\
        \midrule
        Qwen-0.5B-Baseline & 47.5 & 45.8 & 61.3 \\
        \midrule
        \multicolumn{4}{l}{\textbf{Sampling Strategy Choices for tDRO}} \\
        \midrule
         \multicolumn{4}{l}{\textit{Dataset Selection}} \\
         \midrule
         w/ Bottom-50\% & 45.2$\mathrm{{}_{-2.3}}$ & 44.6$\mathrm{{}_{-1.2}}$ & 60.7$\mathrm{{}_{-0.6}}$ \\
         w/ Top-60\% & 48.4$\mathrm{{}_{+0.9}}$ & 48.2$\mathrm{{}_{+2.4}}$ & 62.0$\mathrm{{}_{+0.7}}$ \\
         w/ Top-70\% & \textbf{48.9}$\mathbf{{}_{+1.4}}$ & 48.7$\mathrm{{}_{+2.9}}$ & \textbf{62.3}$\mathbf{{}_{+1.0}}$ \\
         w/ Top-80\% & \textbf{48.9}$\mathbf{{}_{+1.4}}$ & 47.8$\mathrm{{}_{+2.0}}$ & 62.1$\mathrm{{}_{+0.8}}$ \\
         \midrule
         \textit{Samping Ratio Reweighting} & 48.3$\mathrm{{}_{+0.8}}$ & \textbf{49.1}$\mathbf{{}_{+3.3}}$ & 62.2$\mathrm{{}_{+0.9}}$ \\
         \midrule
         \textit{Loss Reweighting} & 48.2$\mathrm{{}_{+0.7}}$ & 46.3$\mathrm{{}_{+0.5}}$ & 61.2$\mathrm{{}_{-0.1}}$ \\
        \toprule
        \multicolumn{4}{l}{\textbf{Loss Measurements (w/ Sampling Ratio Reweighting)}} \\
        \midrule
        $\bar{\mathcal{L}}^{proxy} / \bar{\mathcal{L}}^{ref}$ & \textbf{48.3}$\mathbf{{}_{+0.8}}$ & \textbf{49.1}$\mathbf{{}_{+3.3}}$ & \textbf{62.2}$\mathbf{{}_{+0.9}}$ \\
        $\bar{\mathcal{L}}^{proxy} - \bar{\mathcal{L}}^{ref}$ & 47.5$\mathrm{{}_{-0.0}}$ & 45.3$\mathrm{{}_{-0.5}}$ & 60.1$\mathrm{{}_{+1.2}}$ \\
        $\bar{\mathcal{L}}^{proxy}$ & 47.2$\mathrm{{}_{-0.3}}$ & 47.0$\mathrm{{}_{+1.2}}$& 61.5$\mathrm{{}_{+0.2}}$ \\
    \bottomrule
    \end{tabular}
\caption{Ablation studies on sampling strategy choices and group loss designs.}
\label{table_tdro_ablation}
\end{table}

\subsection{Results}
Retrieval benchmarks on multilingual, cross-lingual, and monolingual English retrieval are conducted and listed in Table \ref{table_main_results}. tDRO optimized domain weights are listed in Figure \ref{fig_weights}. Top-70\% is the optimal line for the top-rated dataset selection strategy. Ablation on the percentages of dataset selection will be presented later.

\subsubsection{tDRO Boosts Retrieval Performances.}
tDRO boosts the universal retrieval performances of LLM-DR on a series of different-sized base LLM, including Qwen-1.5 (0.5B, 1.8B, 4B, 7B), Mistral-0.1 7B, and Llama3 8B. Both the top-rated dataset selection and the sample ratio reweighting strategies work well on most retrieval benchmarks with significant performance gains. The improvements in multilingual MIRACL and cross-lingual MKQA retrieval are attributed to the up-weighting of multilingual datasets, e.g. MIRACL. 

\subsubsection{tDRO Balances Data Usages}
As is shown in Figure \ref{fig_weights}, although multilingual training sets are less than monolingual sets, tDRO balances the lingual distribution differences by top-rating the multilingual MIRACL and T2-Ranking. In contrast, tDRO down-weights 9 monolingual English datasets (Nearly 30\% amounts) to less than 0.01. If we remove less-weighted datasets and keep only the Top-70\% sets, the results on multilingual, cross-lingual, and monolingual retrieval benchmarks even get significantly better, although less data is involved in LLM-DR fine-tuning. This is attributed to the end-to-end task-level data distribution optimization by tDRO, which balances the data distribution and helps LLM-DR training to focus on more useful and challenging data. Notably, as shown in Table \ref{table_sota_scores}, tDRO-optimized LLM-DR gets leading performances without using further enhancements, such as GPT4 data synthesis and special pre-training.


\section{Analysis}

\subsection{Weight Transfering Strategies}
According to Table \ref{table_tdro_ablation}, top-rated dataset selection reaches peak performances with Top-70\% datasets. On the contrary, using Bottom-50\% significantly hurts the performances. This strategy is stable on nearly all benchmarks. However, as is shown in Table \ref{table_main_results}, some monolingual results seem to reach peak levels without further gains, such as Mistral-7B and Llama3-8B for BeIR. This phenomenon only occurs in LLMs with larger parameters ($>$7B) on BeIR. A potential explanation could be that larger LLMs have stronger capacities to tolerate the data weight changes when plenty of monolingual data is provided. Under this circumstance, large LLMs can perform evenly well. 

Sample ratio reweighting has more performance gains on MKQA than the selection strategy because it incurs more sampling probabilities on multilingual datasets, e.g. MIRACL and T2-Ranking, as displayed in Figure \ref{fig_weights}. However, such over-weighting on multilingual datasets is also the reason for no significant gains on monolingual BeIR of large LLMs, e.g. Qwen-4B, 7B, Mistral-7B, and Llama3-8B.

Additionally, we also test re-scaling contrastive losses of LLM-DR with transferred weights. However, such loss reweighting improves BeIR but has no gains on MIRACL and MKQA. A potential reason could be that loss reweighting only changes the loss scale, but does not import more multilingual data because the data sampling ratios are unchanged.

\subsection{Loss Measurement Designs}
Loss measurement is the key design to get proper domain loss scales. tDRO utilizes \textit{relative loss measurement} by dividing the proxy loss with the reference loss. We also tested two additional choices:
\begin{enumerate}
    \item Minus the proxy loss with reference loss. This design is also called group excess loss in DRO \cite{Oren2019DRO}. But obviously, the minus operation could not \textit{scale} the big loss differences of heterogeneous collections for LLM-DR fine-tuning.
    \item Directly using proxy loss.
\end{enumerate}

As is shown in Table \ref{table_tdro_ablation}, these two loss measurements hurt the performances. This is attributed to the incomparable loss scales of different datasets. According to Figure \ref{fig_weights}, both measurements over-prefer Yahoo significantly, because of the biggest loss scale of Yahoo as shown in Table \ref{table_loss_compare_of_qwen500m}.

\section{Related Works}

\subsection{Large Language Model-based Dense Retrieval (LLM-DR)}
Dense retrieval (DR) has gained significant improvements in recent years because of the rapid developments of large language models (LLM) \cite{long2022instructgpt, Hugo2023LLaMA, Bai2024Qwen, Jiang2023Mistral-7B} and growing collections of heterogeneous training data \cite{NilsReimers2021Sentence-Transformers-Training-Data}. In short, the model parameter is growing, and the datasets are increasing. Sentence-T5 \cite{Ni2022SentenceT5} first scales the foundation models to billon parameter T5 and gains good abilities of sentence embeddings. RepLLaMA and RankLLaMA \cite{Ma2024RepLLaMA} first finetune retrievers and re-rankers on Llama2 with a single MS-MARCO dataset and get remarkable improvements over a series of small-sized baseline retrievers. 
E5-Mistral \cite{Wang2024mistral-E5} utilizes 500k GPT3.5/4 synthesized training data and 15 well-collected datasets to fine-tune the retriever, further pushing the boundary of retrieval abilities. It also transfers these data to specially pre-trained small-sized model mE5 \cite{Wang2024mE5} for state-of-the-art (SOTA) multilingual performances. 
However, the utilization of training collection for LLM-DR still relies on intuitional assessments. As discussed in this paper, empirical assigned data choices or sampling ratios, e.g. uniform sampling, incur sub-optimal performances.

\subsection{Distributionally Robust Optimization (DRO)}
Distributionally Robust Optimization (DRO) is an effective way to battle unbalanced data distributions. Topic CVaR \cite{Oren2019DRO}, first proposes to minimize the worst-case loss of each topic. \cite{Sagawa2019GroupDRO} proposes a gradient-based GroupDRO algorithm and successfully improves the worse-group accuracies. DoReMi \cite{Xie2023DoReMi} utilizes GroupDRO in LLM pre-training with minimax optimization for improving perplexities and downstream accuracies. CGD algorithm \cite{Piratla2022CGD-DRO} introduces the inter-group interactions into GroupDRO by substituting the worst-case loss with the inner product score of group gradients. 

Research on DRO for dense retrieval is still minimal. COCO-DR \cite{Yu2022COCO-DR} combines pre-training method coCondenser \cite{gao2022unsupervised} and CGD algorithm to battle the distribution shifts on both the document and query side. However, its scope is limited to small-sized BERT models over clustered groups of one single MS-MARCO \cite{tri2016msmarco} dataset. 
Our study aims to solve the optimization problem of data distribution in LLM-DR fine-tuning, which end-to-end reweights each dataset for optimal performance.

\section{Conclusion}
Large language model-based dense retrieval (LLM-DR) utilizes multiple heterogeneous training datasets in fine-tuning. Previous studies rely on empirical assessments to decide the sampling ratios of each dataset, which incurs unbalanced training data distribution and leads to sub-optimal performances. In our study, we propose a new task-level Distributionally Robust Optimization (tDRO) algorithm to improve the domain generalization ability by end-to-end reweighting the data distribution of each dataset. Experiments on large-scale retrieval benchmarks show steady improvements with less dataset usage.

\bibliography{aaai25}

\clearpage
\section*{Appendix}

\subsection{Training Hyper-parameters}
To facilitate the reproducibility of our work, we arrange all hyper-parameters or 
experiment settings here for tDRO optimization and LLM-DR fine-tuning.

\begin{table}[!h]
    \centering
    \resizebox{\linewidth}{!}{
    \begin{tabular}{l|c|c}
    \toprule
    ~ & \textbf{tDRO} & \textbf{LLM-DR} \\
    \midrule
    Base model & Qwen1.5$_{0.5B}$ & Any LLMs \\
    Representation Pooling & Last token & Last token \\
    Batch size & 2048 & 2048 \\
    Maxlen (query) & 128 & 128 \\
    Maxlen (document) & 512 & 512 \\
    Training steps & 1k & 1k \\
    Warmup steps & 100 & 100 \\
    Model LR $\eta_{\theta}$ & 1e-4 & 1e-4 \\
    Model LR Scheduler & Cosine & Cosine \\
    Model min LR ratio & 0.1 & 0.1 \\
    Model Optimizer & AdamW & AdamW \\
    Weights LR $\eta_{\alpha}$ & 2e-2 & - \\
    Weights LR Scheduler & Constant & - \\
    Half precision & BF16 & BF16 \\
    Negative types & Hard & Hard+in/cross-batch \\
    Contrastive temperature $\tau$ & 0.002 & 0.002 \\
    LoRA $r$ & 8 & 8 \\
    LoRA $\alpha$ & 32 & 32 \\
    LoRA dropout & 0.1 & 0.1 \\
    \bottomrule
    \end{tabular}
    }
\caption{Training hyper-parameters for tDRO optimization and LLM-DR fine-tuning.}
\label{table_hyperparameter}
\end{table}

\subsection{Full Baseline Losses}
The losses of the Qwen1.5-0.5B LLM-DR baseline model are presented below, which is the full version of Table \ref{table_loss_compare_of_qwen500m}. This model is trained with the uniform data sampling ratio for 1k steps with hyper-parameters in Table \ref{table_hyperparameter}. As discussed in the Introduction Section, severe differences in loss scales exist in LLM-DR fine-tuning among different tasks. Thus tDRO proposes to use relative loss measurement for rescaling the running loss scales and derive the proper improving headroom for data distributional optimization.

\begin{table}[!h]
    \centering
    \resizebox{\linewidth}{!}{
    \begin{tabular}{lc|lc}
    \toprule
        \textbf{Task} & \textbf{Loss} & \textbf{Task} & \textbf{Loss} \\
        \midrule
        yahoo\_answers & 3.9257 & hotpotqa & 1.1098 \\
        amazon\_review\_2018 & 2.9479 & codesearchnet & 0.8673 \\
        miracl & 2.2658 & dureader & 0.6925 \\
        nq & 2.1364 & searchQA\_top5\_snippets & 0.6387 \\
        eli5\_question\_answer & 2.1336 & xsum & 0.4839 \\
        medmcqa & 2.1148 & squad\_pairs & 0.2786 \\
        trivia & 1.8961 & AllNLI & 0.2707 \\
        agnews & 1.5894 & quora\_duplicates\_triplets & 0.1549 \\
        t2ranking & 1.5815 & sentence-compression & 0.0496 \\
        stackexchange\_duplicates & 1.3679 & cnn\_dailymail & 0.0465 \\
        msmarco & 1.3312 & SimpleWiki & 0.0248 \\
        mr\_tydi\_combined & 1.3301 & altlex & 0.0163 \\
        gooaq\_pairs & 1.2492 & ~ & ~ \\
    \bottomrule
    \end{tabular}
    }
\caption{Full loss scales of the Qwen1.5-0.5B LLM-DR baseline model.}
\label{table_full_baseline_model}
\end{table}

\subsection{Prompted Embeddings}
Instead of directly encoding the embeddings from textual inputs, our work adds a prompt description before query input to suit different tasks for LLM-DR retrieval. Following \cite{Wang2024mistral-E5}, we did NOT add prompts on the document side to allow us to reuse the pre-built document index across different task scenarios. Such prompted embedding strategy is a common practice for LLM-DR fine-tuning \cite{Wang2024mistral-E5, Su2023Instructor}, which is useful for multitasking. The basic format of the prompted query is: 

Instruct: \textbf{\{\textit{prompt}\}}$\backslash$nQuery: \textbf{\{\textit{query\_text}\}}

\noindent Different prompts for each task are listed as follows. Note that one dataset or task may have multiple prompts for diversity consideration. 

\subsubsection{Prompts for Training Sets}

\begin{enumerate}
    \item \textbf{agnews}: Given a news title, retrieve the news descriptions that match the title
    \item \textbf{AllNLI}(1): Given a premise, retrieve a hypothesis that is entailed by the premise
    \item \textbf{AllNLI}(2): Retrieve semantically similar text.
    \item \textbf{altlex}(1): Given a sentence, retrieve a paraphrase Wikipedia sentence
    \item \textbf{altlex}(2): Given a passage, retrieve a Wikipedia passage that forms paraphrase pairs
    \item \textbf{amazon\_review\_2018}(1): Given a title, retrieve the corresponding reviews from Amazon
    \item \textbf{amazon\_review\_2018}(2): Given a title, retrieve a Amazon review
    \item \textbf{cnn\_dailymail}: Given highlight sentences, retrieve an relevant article that match the sentences
    \item \textbf{codesearchnet}: Given a comment of the function code, retrieve the corresponding code blocks
    \item \textbf{dureader}: Given a Chinese search query, retrieve web passages that answer the question
    \item \textbf{eli5\_question\_answer}: Provided a user question, retrieve the highest voted answers on Reddit ELI5 forum
    \item \textbf{gooaq\_pairs}: Given a web search query, retrieve the corresponding answers from Google
    \item \textbf{hotpotqa}: Given a multi-hop question, retrieve documents that can help answer the question
    \item \textbf{medmcqa}(1): Given a medical query, retrieve relevant passages that answer the query
    \item \textbf{medmcqa}(2): Given a medical question, retrieve passages that answer the question
    \item \textbf{miracl}(1): Given a question, retrieve Wikipedia passages that answer the question
    \item \textbf{miracl}(2): Retrieve Wikipedia passages that answer the question
    \item \textbf{mr\_tydi\_combined}(1): Given a question, retrieve Wikipedia passages that answer the question
    \item \textbf{mr\_tydi\_combined}(2): Retrieve Wikipedia passages that answer the question
    \item \textbf{msmarco}: Given a web search query, retrieve relevant passages that answer the query
    \item \textbf{nq}(1): Given a question, retrieve Wikipedia passages that answer the question
    \item \textbf{nq}(2): Retrieve Wikipedia passages that answer the question
    \item \textbf{quora\_duplicates\_triplets}(1): Given a question, retrieve questions that are semantically equivalent to the given question
    \item \textbf{quora\_duplicates\_triplets}(2): Find questions that have the same meaning as the input question
    \item \textbf{searchQA\_top5\_snippets}(1): Given a question, retrieve text snippets that answer the question
    \item \textbf{searchQA\_top5\_snippets}(2): Retrieve text snippets that answer the question
    \item \textbf{sentence-compression}: Given a sentence, retrieve a short sentence that is semantically equivalent to the given sentence
    \item \textbf{SimpleWiki}(1): Given a Wikipedia sentence, retrieve sentences that are semantically equivalent to the given sentence
    \item \textbf{SimpleWiki}(2): Retrieve semantically similar text.
    \item \textbf{squad\_pairs}(1): Given a question, retrieve Wikipedia passages that answer the question
    \item \textbf{squad\_pairs}(2): Retrieve Wikipedia passages that answer the question
    \item \textbf{stackexchange\_duplicates\_title-body}: Retrieve duplicate questions and passages from StackOverflow forum
    \item \textbf{t2ranking}: Given a Chinese search query, retrieve web passages that answer the question
    \item \textbf{trivia}(1): Given a question, retrieve Wikipedia passages that answer the question
    \item \textbf{trivia}(2): Retrieve Wikipedia passages that answer the question
    \item \textbf{xsum}: Given a news summary, retrieve articles that match the summary
    \item \textbf{yahoo\_answers\_title\_answer}: Given a title, retrieve Yahoo answers that match the title
\end{enumerate}

\subsubsection{Prompts for Multilingual / Cross-lingual Benchmarks}

\begin{enumerate}
    \item \textbf{MIRACL} (18 languages): Given a question, retrieve Wikipedia passages that answer the question
    \item \textbf{MKQA} (25 languages): Given a question, retrieve Wikipedia passages that answer the question
\end{enumerate}

\subsubsection{Prompts for English BeIR Benchmarks}
\ 

Most of the prompts for BeIR are taken from Mistral-E5 \cite{Wang2024mistral-E5} for reproducibility.

\begin{enumerate}
    \item \textbf{ArguAna}: Given a claim, find documents that refute the claim
    \item \textbf{ClimateFEVER}: Given a claim about climate change, retrieve documents that support or refute the claim
    \item \textbf{CQADupStack}: Given a question, retrieve detailed question descriptions from Stackexchange that are duplicates to the given question
    \item \textbf{DBPedia}: Given a query, retrieve relevant entity descriptions from DBPedia
    \item \textbf{FEVER}: Given a claim, retrieve documents that support or refute the claim
    \item \textbf{FiQA2018}: Given a financial question, retrieve user replies that best answer the question
    \item \textbf{HotpotQA}: Given a multi-hop question, retrieve documents that can help answer the question
    \item \textbf{MSMARCO}: Given a web search query, retrieve relevant passages that answer the query
    \item \textbf{NFCorpus}: Given a question, retrieve relevant documents that best answer the question
    \item \textbf{NQ}: Given a question, retrieve Wikipedia passages that answer the question
    \item \textbf{Quora}: Given a question, retrieve questions that are semantically equivalent to the given question
    \item \textbf{SCIDOCS}: Given a scientific paper title, retrieve paper abstracts that are cited by the given paper
    \item \textbf{SciFact}: Given a scientific claim, retrieve documents that support or refute the claim
    \item \textbf{Touche2020}: Given a question, retrieve detailed and persuasive arguments that answer the question
    \item \textbf{TRECCOVID}: Given a query on COVID-19, retrieve documents that answer the query
\end{enumerate}

\subsection{Domain Weights of Different Designs}
Here we list the domain weights of tDRO and other loss measurement designs in Table \ref{table_domain_weights}, which is a detailed version of Figure \ref{fig_weights}.

\begin{table}[!h]
    \centering
    \resizebox{\linewidth}{!}{
    \begin{tabular}{l|c|c|c|c}
    \toprule
        \textbf{Experiments} → & \textbf{Baseline} & \textbf{tDRO} & \multicolumn{2}{c}{\textbf{Other Designs}} \\
        \textbf{Dataset (25 in total)} ↓ & ~ & $\bar{\mathcal{L}}^{proxy} / \bar{\mathcal{L}}^{ref}$ & $\bar{\mathcal{L}}^{proxy} - \bar{\mathcal{L}}^{ref}$ & $\bar{\mathcal{L}}^{proxy}$ \\
        \midrule
        miracl & 0.0400  & 0.1314  & 0.1376  & 0.1193  \\
        medmcqa & 0.0400  & 0.0940  & 0.0886  & 0.0807  \\
        t2ranking & 0.0400  & 0.0885  & 0.0629  & 0.0651  \\
        nq & 0.0400  & 0.0875  & 0.0618  & 0.0615  \\
        eli5 & 0.0400  & 0.0847  & 0.0767  & 0.0807  \\
        yahoo\_answers & 0.0400  & 0.0838  & 0.1537  & 0.1935  \\
        msmarco & 0.0400  & 0.0713  & 0.0546  & 0.0411  \\
        trivia & 0.0400  & 0.0623  & 0.0561  & 0.0532  \\
        gooaq\_pairs & 0.0400  & 0.0600  & 0.0509  & 0.0344  \\
        agnews & 0.0400  & 0.0563  & 0.0541  & 0.0430  \\
        stackexchange\_dups & 0.0400  & 0.0472  & 0.0285  & 0.0242  \\
        amazon\_review\_2018 & 0.0400  & 0.0425  & 0.0750  & 0.1056  \\
        mr\_tydi\_combined & 0.0400  & 0.0252  & 0.0202  & 0.0181  \\
        codesearchnet & 0.0400  & 0.0154  & 0.0129  & 0.0156  \\
        hotpotqa & 0.0400  & 0.0131  & 0.0174  & 0.0186  \\
        dureader & 0.0400  & 0.0110  & 0.0152  & 0.0112  \\
        squad\_pairs & 0.0400  & 0.0072  & 0.0104  & 0.0060  \\
        searchQA\_top5\_snippets & 0.0400  & 0.0066  & 0.0082  & 0.0080  \\
        xsum & 0.0400  & 0.0063  & 0.0063  & 0.0062  \\
        quora\_duplicates & 0.0400  & 0.0019  & 0.0023  & 0.0028  \\
        sentence-compression & 0.0400  & 0.0011  & 0.0022  & 0.0027  \\
        cnn\_dailymail & 0.0400  & 0.0009  & 0.0014  & 0.0025  \\
        SimpleWiki & 0.0400  & 0.0007  & 0.0011  & 0.0023  \\
        altlex & 0.0400  & 0.0006  & 0.0009  & 0.0019  \\
        AllNLI & 0.0400  & 0.0005  & 0.0006  & 0.0016 \\
    \bottomrule
    \end{tabular}
    }
\caption{Detailed domain weights among baseline, tDRO, and other loss measurement designs.}
\label{table_domain_weights}
\end{table}

\subsection{Full Retrieval Performances}
Here we list the full retrieval scores for MIRACL, MKQA, and BeIR in Table \ref{table_miracl}, Table \ref{table_mkqa}, and Table \ref{table_beir}, which is a detailed version of Table \ref{table_main_results}. 


\begin{table*}[!h]
    \centering
    \resizebox{\linewidth}{!}{
    \begin{tabular}{l|cccccccccccccccccc|c|c}
    \toprule
        \textbf{Model} & \textbf{ar} & \textbf{bn} & \textbf{de} & \textbf{en} & \textbf{es} & \textbf{fa} & \textbf{fi} & \textbf{fr} & \textbf{hi} & \textbf{id} & \textbf{ja} & \textbf{ko} & \textbf{ru} & \textbf{sw} & \textbf{te} & \textbf{th} & \textbf{yo} & \textbf{zh} & \textbf{Avg} & \textbf{Gain} \\
        \midrule
        BM25 & 39.5  & 48.2  & 12.0  & 26.7  & 7.7  & 28.7  & 45.8  & 11.5  & 35.0  & 29.7  & 31.2  & 37.1  & 25.6  & 35.1  & 38.3  & 49.1  & 56.1  & 17.5  & 31.9 & ~ \\
        mContriever & 52.5  & 50.1  & 40.8  & 36.4  & 41.8  & 21.5  & 60.2  & 31.4  & 28.6  & 39.2  & 42.4  & 48.3  & 39.1  & 56.0  & 52.8  & 51.7  & 41.5  & 41.0  & 43.1  & ~ \\
        mE5-large-inst & 76.8  & 73.8  & 55.7  & 51.5  & 53.2  & 56.4  & 77.4  & 53.0  & 60.2  & 52.0  & 69.1  & 65.4  & 67.9  & 72.4  & 83.5  & 78.6  & 60.4  & 55.2  & 64.6 & ~  \\
        E5-Mistral & 73.3  & 70.3  & 54.0  & 57.3  & 52.2  & 52.1  & 74.7  & 55.2  & 52.1  & 52.7  & 66.8  & 61.8  & 67.7  & 68.4  & 73.9  & 74.0  & 58.8  & 54.0  & 62.2 & ~  \\
        \midrule
        \multicolumn{20}{l}{\textbf{Uniform Sampling Baselines}}  \\
        \midrule
        Qwen-0.5B & 57.8  & 45.0  & 41.1  & 41.1  & 43.6  & 32.1  & 56.2  & 38.6  & 29.4  & 41.4  & 47.1  & 51.9  & 44.4  & 42.9  & 60.0  & 57.3  & 47.7  & 46.7  & 45.8 & ~  \\
        Qwen-1.8B & 62.0  & 51.9  & 47.2  & 43.2  & 45.9  & 39.9  & 62.4  & 43.4  & 40.9  & 45.0  & 51.1  & 53.6  & 51.3  & 51.5  & 61.9  & 62.6  & 50.9  & 51.4  & 50.9 & ~  \\
        Qwen-4B & 67.7  & 59.8  & 52.1  & 47.5  & 49.5  & 44.8  & 69.3  & 45.1  & 46.3  & 49.0  & 56.3  & 58.7  & 58.5  & 57.7  & 66.7  & 68.0  & 56.2  & 53.5  & 55.9 & ~  \\
        Qwen-7B & 72.2  & 68.6  & 53.6  & 50.5  & 49.6  & 48.8  & 71.8  & 47.3  & 51.7  & 49.7  & 63.0  & 64.4  & 63.0  & 61.9  & 71.9  & 72.1  & 56.6  & 55.8  & 59.6 & ~  \\
        Mistral-7B & 73.5  & 70.5  & 55.6  & 54.0  & 51.2  & 48.2  & 74.6  & 51.7  & 53.5  & 51.6  & 65.0  & 62.6  & 67.5  & 68.8  & 72.8  & 71.8  & 57.5  & 53.6  & 61.3 & ~  \\
        Llama3-8B & 76.6  & 74.1  & 56.3  & 54.1  & 52.0  & 52.6  & 76.7  & 50.2  & 62.7  & 51.9  & 66.8  & 64.4  & 67.4  & 74.7  & 77.9  & 78.0  & 62.4  & 54.6  & 64.1 & ~  \\
        \midrule
        \multicolumn{20}{l}{\textbf{tDRO - Dataset Selection Top-70\%}} \\
        \midrule
        Qwen-0.5B & 60.9  & 47.4  & 43.9  & 46.8  & 45.7  & 33.0  & 59.4  & 42.6  & 29.6  & 43.4  & 51.8  & 54.0  & 48.1  & 46.4  & 64.1  & 60.4  & 48.5  & 50.7  & 48.7\textsuperscript{*} & \textbf{+2.9} \\
        Qwen-1.8B & 66.1  & 57.8  & 48.6  & 47.9  & 46.9  & 42.7  & 65.4  & 45.2  & 43.1  & 47.1  & 56.2  & 55.3  & 54.6  & 56.6  & 68.2  & 66.7  & 52.0  & 53.9  & 54.1\textsuperscript{*} & \textbf{+3.2} \\
        Qwen-4B & 69.8  & 64.3  & 53.9  & 51.4  & 51.3  & 45.8  & 71.4  & 48.5  & 49.3  & 49.9  & 60.9  & 61.0  & 60.7  & 61.7  & 69.3  & 70.6  & 58.8  & 56.7  & 58.6\textsuperscript{*} & \textbf{+2.7} \\
        Qwen-7B & 74.2  & 70.1  & 53.9  & 54.6  & 51.4  & 49.9  & 73.2  & 49.6  & 53.3  & 51.1  & 66.9  & 65.4  & 64.8  & 66.8  & 73.5  & 74.1  & 57.4  & 58.5  & 61.6\textsuperscript{*} & \textbf{+2.0} \\
        Mistral-7B & 74.9  & 74.9  & 55.3  & {58.8}  & 52.7  & 50.7  & 75.6  & 54.2  & 56.8  & 52.6  & 68.7  & 62.8  & 69.1  & 71.3  & 75.6  & 75.4  & 60.9  & 57.3  & 63.8\textsuperscript{*} & \textbf{+2.5} \\
        Llama3-8B & 77.9  & 75.8  & {56.7}  & 58.0  & 55.5  & {55.2}  & {78.1}  & 54.5  & {64.6}  & 53.5  & 71.0  & 66.5  & {70.4}  & 75.1  & {79.8}  & {80.7}  & 63.5  & 57.6  & {66.4}\textsuperscript{*} & \textbf{+2.3} \\
        \midrule
        \multicolumn{20}{l}{\textbf{tDRO - Sample Ratio Reweighting}} \\
        \midrule
        Qwen-0.5B & 60.8  & 49.9  & 44.5  & 47.6  & 46.8  & 33.9  & 58.2  & 43.9  & 29.2  & 43.4  & 52.4  & 54.3  & 48.3  & 45.6  & 62.0  & 60.0  & 49.5  & 53.1  & 49.1\textsuperscript{*} & \textbf{+3.3} \\
        Qwen-1.8B & 64.9  & 58.0  & 49.7  & 47.3  & 47.7  & 41.8  & 64.0  & 47.1  & 40.2  & 46.1  & 56.4  & 55.1  & 53.9  & 54.4  & 65.7  & 65.4  & 51.7  & 56.0  & 53.6\textsuperscript{*} & \textbf{+2.7} \\
        Qwen-4B & 69.7  & 63.8  & 53.8  & 52.1  & 51.5  & 45.1  & 70.0  & 50.4  & 48.2  & 50.1  & 61.1  & 59.9  & 61.2  & 60.0  & 68.5  & 70.9  & 57.0  & 58.3  & 58.4\textsuperscript{*} & \textbf{+2.5} \\
        Qwen-7B & 73.7  & 69.9  & 53.6  & 53.7  & 51.4  & 49.3  & 71.4  & 50.4  & 51.1  & 50.2  & 65.4  & 64.7  & 64.5  & 64.0  & 73.3  & 74.0  & 58.2  & 58.6  & 61.0\textsuperscript{*} & \textbf{+1.4} \\
        Mistral-7B & 75.4  & 72.5  & 55.0  & 57.7  & 53.4  & 51.1  & 74.4  & 53.3  & 56.5  & 52.7  & 67.1  & 63.8  & 68.9  & 71.4  & 74.2  & 74.3  & 61.0  & 58.7  & 63.4\textsuperscript{*} & \textbf{+2.1} \\
        Llama3-8B & {78.2}  & {75.9}  & 56.0  & 58.3  & {54.5}  & 55.1  & 77.2  & 54.6  & 62.7  & {54.1}  & {72.0}  & {67.8}  & 68.6  & {75.7}  & 78.5  & 80.5  & {64.3}  & {59.0}  & 66.3\textsuperscript{*} & \textbf{+2.2} \\
    \bottomrule
    \end{tabular}
    }
\caption{Multilingual retrieval performance on MIRACL dev sets with 18 languages (measured by nDCG@10). \textsuperscript{*}Significant improvements (p $\leq$ 0.01) over the corresponding baseline.}
\label{table_miracl}
\end{table*}

\begin{table*}[!h]
    \centering
    \resizebox{\linewidth}{!}{
    \begin{tabular}{l|ccccccccccccccccccccccccc|c|c}
    \toprule
    \textbf{Model} & \textbf{ar} & \textbf{da} & \textbf{de} & \textbf{es} & \textbf{fi} & \textbf{fr} & \textbf{he} & \textbf{hu} & \textbf{it} & \textbf{ja} & \textbf{km} & \textbf{ko} & \textbf{ms} & \textbf{nl} & \textbf{no} & \textbf{pl} & \textbf{pt} & \textbf{ru} & \textbf{sv} & \textbf{th} & \textbf{tr} & \textbf{vi} & \textbf{zh\_cn} & \textbf{zh\_hk} & \textbf{zh\_tw} & \textbf{Avg} & \textbf{Gain} \\
    \toprule
        BM25 & 18.9  & 49.3  & 35.4  & 43.4  & 46.3  & 45.3  & 26.9  & 38.2  & 45.2  & 24.5  & 27.8  & 27.9  & 55.9  & 56.2  & 52.1  & 40.8  & 44.9  & 33.2  & 54.6  & 37.8  & 45.8  & 46.6  & 31.0  & 35.0  & 33.5  & 39.9  & ~ \\
        mContriever & 58.2  & 73.9  & 71.7  & 72.6  & 70.2  & 72.8  & 63.8  & 69.7  & 72.3  & 64.8  & 26.8  & 59.7  & 74.1  & 73.7  & 73.5  & 71.6  & 72.0  & 69.8  & 73.2  & 66.9  & 71.1  & 70.9  & 68.1  & 68.0  & 67.9  & 67.9  & ~ \\
        mE5-large-inst & 66.2  & 76.5  & 76.0  & 75.6  & 72.3  & 76.7  & 61.3  & 73.8  & 76.5  & 61.9  & 44.9  & 46.9  & 76.1  & 77.8  & 76.3  & 76.4  & 76.5  & 75.9  & 77.0  & 70.8  & 74.1  & 75.6  & 64.6  & 62.8  & 61.9  & 70.2  & ~ \\
        E5-Mistral-7b & 56.4  & 76.9  & 76.0  & 76.9  & 70.0  & 77.3  & 44.1  & 74.0  & 76.5  & 63.9  & 31.2  & 57.3  & 75.9  & 78.3  & 75.9  & 75.8  & 76.7  & 73.7  & 77.4  & 63.8  & 71.4  & 70.0  & 67.4  & 62.9  & 64.3  & 68.6  & ~ \\
        \midrule
        \multicolumn{28}{l}{\textbf{Uniform Sampling Baselines}} \\
        \midrule
        Qwen-0.5B & 47.4  & 66.2  & 66.7  & 67.8  & 54.7  & 70.1  & 46.8  & 54.0  & 65.4  & 58.1  & 38.3  & 47.8  & 66.2  & 69.8  & 64.1  & 61.8  & 66.3  & 61.8  & 67.2  & 58.8  & 57.8  & 65.2  & 71.0  & 70.0  & 69.6  & 61.3  & ~ \\
        Qwen-1.8B & 52.4  & 69.1  & 69.1  & 70.3  & 59.5  & 71.0  & 48.8  & 58.7  & 67.4  & 59.5  & 39.8  & 53.9  & 69.3  & 72.8  & 66.7  & 65.6  & 68.4  & 65.6  & 69.7  & 60.7  & 61.6  & 65.0  & 72.7  & 71.1  & 71.8  & 64.0  & ~ \\
        Qwen-4B & 60.0  & 75.0  & 73.3  & 74.9  & 68.6  & 75.6  & 58.0  & 68.9  & 74.5  & 70.7  & 48.1  & 54.7  & 73.7  & 76.4  & 73.6  & 72.7  & 74.3  & 73.0  & 74.8  & 67.4  & 69.1  & 72.4  & 75.5  & 74.3  & 75.4  & 70.2  & ~ \\
        Qwen-7B & 66.7  & 76.8  & 76.1  & 76.7  & 71.4  & 77.6  & 64.3  & 72.4  & 75.6  & 74.3  & 52.4  & 67.0  & 75.8  & 78.4  & 75.3  & 75.1  & 76.8  & 76.6  & 76.8  & 73.5  & 72.2  & 75.8  & 77.5  & 76.3  & 77.6  & 73.6  & ~ \\
        Mistral-7B & 63.6  & 78.1  & 77.1  & 78.0  & 71.8  & 78.3  & 57.5  & 75.8  & 76.9  & 73.6  & 39.5  & 68.7  & 77.4  & 78.9  & 76.5  & 77.2  & 77.4  & 77.3  & 78.1  & 70.7  & 73.7  & 72.8  & 74.8  & 72.8  & 73.1  & 72.8  & ~ \\
        Llama3-8B & 72.1  & 78.1  & 77.8  & 78.1  & 76.3  & 78.4  & 73.5  & 76.0  & 77.3  & 77.1  & 50.8  & 72.1  & \textbf{78.6}  & 79.2  & 77.3  & 77.8  & 78.1  & 78.0  & 78.3  & 77.6  & 76.7  & 77.3  & 76.7  & 75.5  & 76.5  & 75.8  & ~ \\
        \midrule
        \multicolumn{28}{l}{\textbf{tDRO - Dataset Selection Top-70\%}} \\
        \midrule
        Qwen-0.5B & 48.3  & 67.4  & 68.0  & 69.1  & 55.4  & 70.7  & 48.2  & 54.3  & 66.9  & 59.3  & 38.9  & 49.1  & 67.1  & 71.2  & 65.5  & 63.1  & 67.3  & 63.2  & 68.6  & 59.4  & 58.7  & 65.9  & 71.9  & 70.8  & 70.3  & 62.3\textsuperscript{*}  & \textbf{+1.0}  \\
        Qwen-1.8B & 54.4  & 70.1  & 70.3  & 71.5  & 61.6  & 72.3  & 50.6  & 60.8  & 69.6  & 62.5  & 43.0  & 54.7  & 69.8  & 72.9  & 68.2  & 67.2  & 70.2  & 66.8  & 71.6  & 62.8  & 62.9  & 66.7  & 73.8  & 72.7  & 72.9  & 65.6\textsuperscript{*}  & \textbf{+1.6}  \\
        Qwen-4B & 63.3  & 75.5  & 74.2  & 75.9  & 69.3  & 76.1  & 62.0  & 70.2  & 75.5  & 70.8  & 47.6  & 61.6  & 74.4  & 76.8  & 74.4  & 73.6  & 75.1  & 74.5  & 75.7  & 69.2  & 70.0  & 73.7  & 76.0  & 74.8  & 75.3  & 71.4\textsuperscript{*}  & \textbf{+1.2}  \\
        Qwen-7B & 67.7  & 77.2  & 76.4  & 76.8  & 71.5  & 77.6  & 63.4  & 73.2  & 76.2  & 74.7  & 52.8  & 65.3  & 75.8  & 78.5  & 75.7  & 75.4  & 77.1  & 76.7  & 77.3  & 74.3  & 72.6  & 76.4  & \textbf{77.7}  & 76.7  & 77.8  & 73.8  & +0.2  \\
        Mistral-7B & 65.6  & 78.3  & 77.8  & 78.3  & 72.9  & 78.4  & 61.4  & 76.2  & 77.6  & 75.5  & 42.9  & 71.7  & 77.5  & 78.8  & 77.0  & 77.5  & 77.6  & 77.9  & 78.3  & 72.2  & 74.2  & 74.0  & 75.8  & 73.8  & 74.1  & 73.8\textsuperscript{*}  & \textbf{+1.0}  \\
        Llama3-8B & 73.0  & 78.3  & 78.3  & 78.4  & 76.1  & 78.7  & 74.1  & 76.5  & 77.9  & 77.4  & 57.8  & 73.5  & 78.3  & 79.3  & 77.6  & 77.9  & 78.1  & 78.0  & 78.4  & 78.6  & 77.0  & 77.7  & 77.4  & 76.1  & 76.8  & 76.4\textsuperscript{*}  & \textbf{+0.6}  \\
        \midrule
        \multicolumn{28}{l}{\textbf{tDRO - Sample Ratio Reweighting}} \\
        \midrule
        Qwen-0.5B & 48.6  & 67.4  & 67.3  & 68.9  & 55.0  & 70.6  & 47.9  & 53.3  & 66.4  & 59.9  & 39.1  & 48.8  & 67.3  & 70.9  & 65.1  & 63.1  & 67.8  & 63.1  & 68.2  & 59.9  & 58.4  & 65.4  & 71.8  & 71.0  & 70.5  & 62.2\textsuperscript{*}  & \textbf{+0.9}  \\
        Qwen-1.8B & 55.3  & 71.8  & 71.7  & 70.8  & 62.0  & 72.6  & 52.8  & 61.5  & 71.7  & 66.1  & 43.8  & 56.7  & 70.3  & 74.4  & 69.6  & 67.7  & 71.0  & 68.0  & 72.6  & 64.2  & 63.8  & 69.2  & 75.0  & 73.8  & 74.3  & 66.8\textsuperscript{*}  & \textbf{+2.8}  \\
        Qwen-4B & 63.5  & 76.2  & 75.4  & 76.5  & 70.4  & 76.6  & 62.1  & 70.5  & 75.9  & 73.1  & 48.2  & 60.8  & 74.1  & 77.5  & 74.7  & 73.9  & 76.1  & 75.0  & 76.4  & 69.9  & 70.5  & 73.5  & 77.1  & 75.9  & 76.9  & 72.0\textsuperscript{*}  & \textbf{+1.8}  \\
        Qwen-7B & 66.9  & 77.1  & 76.8  & 76.5  & 71.3  & 77.5  & 61.8  & 72.3  & 76.1  & 75.3  & 53.7  & 64.9  & 75.3  & 78.5  & 75.7  & 75.4  & 76.7  & 76.9  & 77.0  & 73.9  & 72.3  & 75.8  & 77.9  & \textbf{77.0}  & \textbf{78.1}  & 73.6  & +0.0  \\
        Mistral-7B & 65.3  & 78.3  & 77.8  & \textbf{78.6}  & 72.8  & 78.7  & 61.3  & 76.5  & 77.9  & 75.6  & 43.6  & 71.4  & 77.5  & 79.1  & 77.2  & \textbf{78.1}  & 77.8  & 78.1  & \textbf{78.6}  & 72.3  & 74.4  & 74.0  & 75.7  & 73.7  & 74.6  & 74.0\textsuperscript{*}  & \textbf{+1.2}  \\
        Llama3-8B & \textbf{74.2}  & \textbf{78.5}  & \textbf{78.4}  & 78.4  & \textbf{76.6}  & \textbf{78.9}  & \textbf{74.5}  & \textbf{76.8}  & \textbf{78.5}  & \textbf{78.0}  & \textbf{59.2}  & \textbf{74.4}  & 78.5  & \textbf{79.5}  & \textbf{77.7}  & 77.9  & \textbf{78.4}  & \textbf{78.6}  & \textbf{78.6}  & \textbf{78.8}  & \textbf{77.1}  & \textbf{77.8}  & \textbf{77.7}  & 76.6  & 77.2  & \textbf{76.8}\textsuperscript{*}  & \textbf{+1.0} \\
    \bottomrule
    \end{tabular}
    }
\caption{Cross-lingual retrieval performance on MKQA test sets with 25 languages (measured by Accuacy@100).}
\label{table_mkqa}
\end{table*}

\begin{table*}[!h]
    \centering
    \resizebox{\linewidth}{!}{
    \begin{tabular}{l|ccccccccccccccc|c|c}
    \toprule
        \textbf{Model} & \textbf{ArguAna} & \textbf{CQADup} & \textbf{C-FEVER} & \textbf{DBPedia} & \textbf{FEVER} & \textbf{FiQA} & \textbf{HotpotQA} & \textbf{MSMARCO} & \textbf{NFCorpus} & \textbf{NQ} & \textbf{Quora} & \textbf{SCIDOCS} & \textbf{SciFact} & \textbf{T-COVID} & \textbf{Touche} & \textbf{Avg} & \textbf{Gain} \\
        \midrule
        BM25 & 31.5  & 29.9  & 21.3  & 31.3  & 75.3  & 23.6  & 60.3  & 22.8  & 32.5  & 32.9  & 78.9  & 15.8  & 66.5  & 65.6  & \textbf{36.7}  & 41.7  & ~ \\
        Contriever & 44.6  & 34.5  & 23.7  & 41.3  & 75.8  & 32.9  & 63.8  & 36.8  & 32.8  & 49.8  & 86.5  & 16.5  & 67.7  & 59.6  & 23.0  & 46.0  & ~ \\
        mE5-large-inst & 55.5  & 42.7  & \textbf{29.8}  & 38.4  & 78.0  & 47.7  & 69.3  & 40.4  & 35.6  & 57.7  & 89.1  & 18.7  & 71.9  & \textbf{82.0}  & 27.3  & 52.3  & ~ \\
        E5-Mistral-7b$\dagger$ & \textbf{62.5}  & 42.9  & 25.2  & \textbf{47.7}  & 73.1  & 54.5  & 75.6  & \textbf{42.9}  & 35.3  & 57.3  & \textbf{89.5}  & 19.0  & 74.7  & 70.8  & 19.1  & 52.7  & ~ \\
        \midrule
        \multicolumn{18}{l}{\textbf{Uniform Sampling Baselines}} \\
        \midrule
        Qwen-0.5B & 55.4  & 41.5  & 24.4  & 34.8  & 64.3  & 39.3  & 62.2  & 32.5  & 34.4  & 47.5  & 87.7  & 18.5  & 68.9  & 73.9  & 26.6  & 47.5  & ~ \\
        Qwen-1.8B & 54.6  & 44.1  & 25.6  & 36.6  & 69.3  & 44.3  & 65.0  & 32.9  & 36.9  & 47.8  & 87.6  & 19.8  & 70.5  & 75.2  & 22.4  & 48.8  & ~ \\
        Qwen-4B & 57.9  & 47.7  & 26.1  & 41.3  & 75.3  & 49.6  & 70.6  & 35.9  & 38.2  & 53.3  & 87.9  & 22.0  & 74.3  & 77.2  & 19.7  & 51.8  & ~ \\
        Qwen-7B & 58.4  & 49.8  & 25.7  & 41.3  & 75.0  & 51.5  & 72.7  & 36.1  & 39.1  & 55.0  & 88.5  & 22.3  & 76.0  & 73.5  & 19.7  & 52.3  & ~ \\
        Mistral-7B & 59.6  & 50.6  & 26.6  & 45.2  & 78.5  & 58.6  & 78.8  & 39.0  & \textbf{42.0 } & 60.6  & 88.7  & 22.3  & \textbf{79.2}  & 77.9  & 20.4  & 55.2  & ~ \\
        Llama3-8B & 58.2  & 51.3  & 27.0  & 44.2  & 79.0  & 57.7  & 78.9  & 38.5  & 41.3  & 60.6  & 89.0  & \textbf{22.9}  & 78.6  & 75.8  & 21.3  & 55.0  & ~ \\
        \midrule
        \multicolumn{18}{l}{\textbf{tDRO - Dataset Selection Top-70\%}} \\
        \midrule
        Qwen-0.5B & 55.2  & 41.8  & 25.2  & 37.8  & 72.2  & 41.7  & 63.9  & 33.4  & 35.6  & 49.4  & 84.8  & 18.7  & 69.4  & 76.9  & 27.3  & 48.9\textsuperscript{*}  & \textbf{+1.4}  \\
        Qwen-1.8B & 56.8  & 44.8  & 26.4  & 37.4  & 76.8  & 46.0  & 67.0  & 33.0  & 37.1  & 51.9  & 86.8  & 19.9  & 72.5  & 73.9  & 22.8  & 50.2\textsuperscript{*}  & \textbf{+1.4}  \\
        Qwen-4B & 59.0  & 46.9  & 26.7  & 41.4  & 80.1  & 51.1  & 72.2  & 36.3  & 38.6  & 56.3  & 87.7  & 21.6  & 75.2  & 75.8  & 19.6  & 52.6\textsuperscript{*}  & \textbf{+0.8}  \\
        Qwen-7B & 59.2  & 49.6  & 25.5  & 40.9  & 77.8  & 53.6  & 73.9  & 36.0  & 40.0  & 57.8  & 87.2  & 22.3  & 77.2  & 75.6  & 22.2  & 53.3\textsuperscript{*}  & \textbf{+1.0}  \\
        Mistral-7B & 53.3  & 50.3  & 25.5  & 44.4  & 81.9  & 59.7  & 79.1  & 39.7  & 41.9  & 63.4  & 87.0  & 21.4  & 77.1  & 79.4  & 23.9  & \textbf{55.2}  & +0.0 \\
        Llama3-8B & 56.9  & \textbf{51.5}  & 28.1  & 45.0  & 81.1  & 57.8  & \textbf{80.0}  & 38.9  & 41.6  & 61.9  & 87.8  & 22.1  & 78.2  & 75.7  & 19.7  & 55.1  & +0.1  \\
        \midrule
        \multicolumn{18}{l}{\textbf{tDRO - Sample Ratio Reweighting}} \\
        \midrule
        Qwen-0.5B & 55.6  & 40.5  & 22.0  & 37.0  & 70.5  & 42.2  & 60.9  & 34.2  & 34.9  & 51.1  & 84.5  & 18.0  & 69.7  & 78.0  & 25.2  & 48.3\textsuperscript{*}  & \textbf{+0.8}  \\
        Qwen-1.8B & 57.2  & 44.1  & 23.6  & 37.3  & 73.0  & 47.0  & 62.3  & 34.1  & 37.5  & 53.1  & 86.3  & 19.5  & 72.9  & 74.0  & 23.0  & 49.7\textsuperscript{*}  & \textbf{+0.9}  \\
        Qwen-4B & 59.9  & 46.3  & 24.4  & 41.5  & 76.1  & 51.4  & 67.5  & 36.8  & 38.8  & 57.5  & 87.4  & 21.7  & 76.4  & 73.7  & 19.1  & 51.9  & +0.1  \\
        Qwen-7B & 57.4  & 48.6  & 24.7  & 40.8  & 76.4  & 51.6  & 69.3  & 37.3  & 39.9  & 59.4  & 87.2  & 21.7  & 77.4  & 75.9  & 18.5  & 52.4  & +0.1  \\
        Mistral-7B & 55.6  & 49.1  & 25.7  & 44.1  & 80.1  & \textbf{60.6}  & 75.6  & 41.1  & 41.5  & \textbf{63.2}  & 86.0  & 21.9  & 78.6  & 81.6  & 25.9  & \textbf{55.4}  & +0.2  \\
        Llama3-8B & 51.9  & 50.6  & 28.9  & 42.9  & \textbf{81.7}  & 56.9  & 75.8  & 39.9  & 41.2  & 63.1  & 87.9  & 22.2  & 78.7  & 80.5  & 22.5  & 55.0  & +0.0 \\
    \bottomrule
    \end{tabular}
    }
\caption{English retrieval performance on BeIR test sets (except MS-MARCO, which uses dev set.) with 15 datasets (measured by nDCG@10). $\dagger$E5-Mistral-7b scores without using GPT synthesis data are listed here for fairness consideration.}
\label{table_beir}
\end{table*}


\end{document}